\documentclass[preprint,aps,floatfix,showpacs,a4paper]{revtex4}

\usepackage{amsmath,amssymb,amsfonts,dcolumn,color,graphicx,graphics,latexsym,placeins,epsfig}

\newcommand{\be}{\begin{equation}}
\newcommand{\ee}{\end{equation}}
\newcommand{\ba}{\begin{eqnarray}}
\newcommand{\ea}{\end{eqnarray}}

\newcommand{\p}{\prime}
\newcommand{\4}{\frac}

\newcommand{\la}{\label}

\renewcommand{\inf}{\infty}

\begin{document}

\title{\Large \bf The Ellis wormhole with `tachyon matter'}

\author{A. Das}
\email{anupam@phy.iitkgp.ernet.in}
\affiliation{\rm Department of Physics and Meteorology, Indian Institute of Technology, Kharagpur, 721302, India}

\author{Sayan Kar}
\email{sayan@cts.iitkgp.ernet.in}
\affiliation{\rm Department of Physics and Meteorology {\it and} Center for Theoretical Studies \\Indian Institute of Technology, Kharagpur, 721302, India}

\begin{abstract}
The Ellis wormhole is known to be an exact solution of the Einstein--scalar
system where the scalar field has negative kinetic energy (phantom). 
We show here that the same geometry (in $3+1$ dimensions) 
can also be obtained with `tachyon matter' 
as a source term in the field equations and a positive cosmological constant. 
The kink--like tachyon field profile and the nature of the tachyon potential 
are analyzed. It is shown that the field configuration and the spacetime are 
stable against linear scalar perturbations. Finally, we comment on extensions 
of the $3+1$ dimensional Ellis wormhole (with tachyon matter source) 
in diverse dimensions ($d=2+1$ and $d>4$).
\end{abstract}

\pacs{04.20.-q, 04.20.Jb}

\maketitle

\noindent{\sf Introduction :} The Ellis geometry obtained in 1973 {\cite{ellis1}}, 
is the first example of a nonsingular wormhole solution. It makes use of
a wrong--sign kinetic energy term  (the so--called `phantom' of today)
in the scalar field action to achieve
the crucial energy condition violation, so necessary to support a
wormhole (recent works showing construction of womholes with phantom matter are {\cite{sushkov,lobo}}). Post 1988, 
the wormhole industry got a boost through the
papers of Morris, Thorne, Visser and others {\cite{mt1}}. With the
possibility of time--machine models using wormholes {\cite{mt2, frolov, visser_t}} and the
issue of energy condition violations, wormholes of Lorentzian
signature became an area of active research. Very recently, issues of stability and time-travel using wormholes have been discussed afresh {\cite{stab}}. For a summary of wormholery
till 1995, see the book by Visser{\cite{vissbook}} and for more uptodate account of developments in wormhole research, consult {\cite{lemos}} and the references therein. More recently,
the notion of quantifying energy condition violations 
with appropriate examples have been dealt with in {\cite{skmvnd1,skmvnd2}.

What is special about Ellis geometry apart from the fact that it is
wormhole? Firstly, the $\theta=\frac{\pi}{2}$ two dimensional section
of the spacelike hypersurface of this geometry when embedded in three
dimensional Euclidean space is the well--known `catenoid'--a minimal
surface. Secondly, the geodesic equations as well as wave propagation in this 
geometry are exactly solvable -- in particular, scalar waves in Ellis geometry 
(and its generalizations) have been analyzed in {\cite{sk1}}.
Thirdly, amidst the lack of viable sources for wormholes, 
the geometry provides a `solution' to the
full Einstein--scalar system with negative scalar kinetic energy.     
We also mention here that the Ellis geometry can be
obtained with a Kalb--Ramond axion as a source (see {\cite{skssg}} for
further details).

In this article, we show that the same Ellis geometry can be obtained as a 
solution to the Einstein equations with tachyon matter as the source
and in the presence of a positive cosmological constant. We shall see how the
cosmological constant is crucial in obtaining this solution.
The tachyon field profile resembles a kink and the tachyon potential
is well--like. Furthermore, a study of perturbations about the scalar field and 
the metric is carried out in order to check the stability of our
solution. We also comment on how the geometry may be generalized
with a non--trivial redshift function (ie, $g_{tt} \neq -1$).  Finally, 
we show how the tachyon field can act as source for wormhole solutions in 
lower ($2+1$) as well as higher ($D>4$) dimensions. We will work in a proper orthonormal basis. Our metric signature is $(-,+,+,+)$ and its extension to higher dimensions. We choose $c=1$ unless otherwise stated.

\noindent{\sf Action, equations of motion and solutions:} 
We begin with the general form  of a static spherically 
symmetric line element in $(3+1)$ dimensions:
\be
ds^2 = -{{\psi}^2}(l) dt^2 + dl^2 + r^2(l)(d\theta^2 + \sin^2\theta d\phi^2) \la{metric}
\ee
where $\psi(l)$ is the redshift function and $r(l)$ is a function which 
determines the shape of the wormhole. 
Here $l$ corresponds to the proper radial distance of an event in the 
$(t,l,\theta,\phi)$ coordinate system. As usual, the ranges of the coordinates are:$\, -\inf < t < \inf,-\inf < l < \inf, 0\leq \theta \leq \pi, 0\leq  \phi \leq 2 \pi $. Note that for the geometry to
represent a wormhole, $(i)$ The shape function 
must have a global minimum (say at $r(l)|_{l=0}=r_0)$ which connects two asymptotically flat regions (located at $l \rightarrow \pm \infty$). This implies that  $r'(l=0)=0$ and $r''(l=0)>0$ (prime denotes derivative of the function). $(ii)$ To satisfy the criterion of asymptotic flatness stated above, $Lim_{l\rightarrow \pm \infty} r(l)/|l| = 1$. $(iii)$ In addition, to avoid the existence of horizons $\psi(l)$ must be nonzero for all values of proper radial distance. This condition is required to get a wormhole `traversable in principle'.
 
The background matter constituting the wormhole is assumed to be the 
newly proposed 
`tachyon matter' (tachyon condensate) represented by a scalar field $T$. Its origin lies in string 
theory, where the presence of tachyonic modes of open strings is related to the instability 
of D-branes. Analysis has shown that in 
the low energy effective field theory for the tachyon condensate, 
the corresponding tachyon potential must have an unstable maximum at the 
origin of the field, which can be shown to be consistent with the fact that 
tachyons have negative $(mass)^2$ (consult {\cite{sen2}} for details). 
The tachyon tends to roll down the maximum 
of potential and evolve to a stable vacuum expectation value.
This `rolling tachyon' has been made use of extensively in cosmology
(with a purely time--dependent tachyon) {\cite{kim2}}. Spatially dependent tachyon
configurations have been studied in different contexts with varying
tachyon potentials (for discussion and references consult {\cite{gorini, dao, kim}}). However, the kind of tachyon 
configuration in a curved nonsingular background obtained here in this article is, 
as of now, not discussed in the context of string theory. Thus, we
may take the attitude of assuming the tachyon condensate action 
as a new type of scalar field theory coupled to gravity in which
we search for exact solutions.

The low energy effective action for the tachyon scalar coupled to
gravity is given by  \cite{sen1, gibbons1}:
\be
S\!=\!\!\!\int\!\!\sqrt{-g} d^4x \left( \frac{R}{2k} + V(T)(1+T^{,\mu} 
T_{,\mu})^{1/2}\right) \la{action}
\ee
where $R$ is the Ricci scalar for the spacetime defined as 
and $k=8 \pi G$ is a constant.  $V(T)$ is an 
unspecified function of the field and is known as the tachyon potential. 
The tachyon field couples to the system in a nontrivial 
way.
The stress-energy tensor of the tachyon field obtained by varying the 
matter part of the action is 
\be
T_{\mu\nu} = V(T)[g_{\mu\nu}(1+T^{,\alpha} T_{,\alpha})^{1/2} - \frac{T_{,\mu} T_{,\nu}}{(1+T^{,\alpha}T_{,\alpha})^{1/2}}]
\ee

Assuming a background line element of the form given in (1) and with tachyon field a 
function of the proper radial distance only, the nonzero components of the energy 
momentum tensor in an orthogonal basis are as follows:
\begin{eqnarray}
T_{tt}&=& - V(T)(1 + T^{\p 2})^{1/2} = \rho(T) \label{Tt}\\
T_{rr}&=& \4{V(T)}{(1+{T}^{\p 2})^{1/2}} = p_r(T) \label{Tr}\\
T_{\theta\theta}&=& V(T)(1 + T^{\p 2})^{1/2} = p_{tr}(T) \label{Tth}
\end{eqnarray}
where the components $T_{tt}, T_{rr}, T_{\theta\theta}$ are identified as the energy density, radial and transverse pressure of the tachyon field respectively. 
The components of the Einstein tensor for the line element in (1) are:
\ba
G_{tt}&=& -\4{2r^{\p \p}}{r} - (\4{r^{\p}}{r})^2 +\4{1}{r^2}\\
G_{rr}&=& \4{2\psi^{\p} r^{\p}}{\psi r} + (\4{r^{\p}}{r})^2 -\4{1}{r^2}\\
G_{\theta\theta}&=& \4{\psi^{\p\p}}{\psi}+ \4{\psi^{\p} r^{\p}}{\psi r}+\4{r^{\p \p}}{r}
\ea
where the prime denotes derivative of the function with respect to proper radial distance. We will first check possibility of getting consistent solutions for the unknown 
functions $\psi(l), r(l), T(l)$ and $V(T)$ by solving the Einstein field 
equations without a $\Lambda$ term (i.e. $G_{\mu\nu} = kT_{\mu\nu}$). 
Note from (4) and (6) that $T_{tt}=-T_{\theta\theta}$. Therefore, assuming the Einstein
field equations to hold good we must have $G_{tt}= -G_{\theta\theta}$. This relation is crucial and will be repeatedly used in this report. This leads to the condition
\be
\4{r^{\p \p}}{r} + (\4{r^{\p}}{r})^2 -\4{1}{r^2} - \4{r^{\p}\psi^{\p}}{r\psi} -\4{\psi^{\p\p}}{\psi}=0 \label{cons}
\ee
It may be noted here that this condition does not change even if we
include a non--zero cosmological constant.

It is obvious that solutions for all the unknown functions in the problem 
cannot be obtained without suitable input on at least one of them. 
Assuming $\psi(l)=\xi(l) \eta (l) = \frac{\eta(l)}{\sqrt{r(l)}}$ and 
substituting it in (10) 
we get
\be
\eta^{\p\p}+\eta \left(\4{\xi^{\p\p}}{\xi}+\4{\xi^{\p}r^{\p}}{\xi r}-\4{r^{\p \p}}{r}-(\4{r^{\p}}{r})^2+\4{1}{r^2}\right)=0
\ee 

Choosing $r(l)=(b_0^2+l^2)^{1/2}$ (Ellis geometry, $b_0$ being a constant) 
the above equation (\ref{cons}) becomes,
\be
\eta^{\p\p}+\left( \4{-2b_0^2 + l^2}{4(b_0^2+l^2)^2}\right)\eta=0
\ee
with solutions given by :
\ba
\eta_1(l)&=&C_2(b_0^2+l^2)^{1/4} \implies \psi_1(l)=\mathrm{Constant} \\
\eta_2(l)&=&C_3(b_0^2+l^2)^{1/4}\sinh^{-1}{(l/b_0)}\implies\psi_2(l)=C_3\sinh^{-1}{(l/b_0)}
\ea

Thus (i) if the shape function is to be that of an Ellis geometry and (ii) the
criterion $G_{tt}=-G_{\theta\theta}$ has to hold good (requirement for a
tachyon matter source) then we are left with the above two
choices for the red-shift function.

The second solution for $\psi(l)$ can be shown to give rise to an unphysical 
tachyon field. Using $r(l)$, $\psi(l)$ in the expressions for $G_{tt}$, 
$G_{rr}$ and the Einstein equations, we find that ${T'}^2(l)$ is not a positive definite function, which is not desirable and we discard this solution.
 We therefore have to work with a constant red-shift function which can be chosen to have the value unity.

Note that $r(l)$ satisfies conditions on a typical wormhole shape 
function: i.e., presence of a global minimum (at $l=0$, throat of the wormhole) and the condition of asymptotic flatness. With our choice of constant redshift function 
and the solution for $r(l)$ the $(tt)$ and $(rr)$ components of the Einstein 
tensor simplify to 
\be
G_{tt} = G_{rr} = -\4{r^{\p \p}}{r}\\
\ee
However, we notice that the relations  $G_{tt}/G_{rr} = T_{tt}/T_{rr}$ and  
$G_{tt}G_{rr} = k^2 T_{tt}T_{rr}$ lead respectively to  
\ba
T^{\p}(l)&=& 0\\
(\4{r^{\p \p}}{r})^2&=& - k^2 V(T)^2
\ea
which indicates that the function $V(T)$ has to be imaginary and the tachyon 
field is a constant. 

How do we circumvent the above problem? We now show that that inclusion of 
the $\Lambda$ term in the field equations can give rise to consistent solutions for the field and the potential.
The field equations with $\Lambda$ term become
\ba
G_{ij} + \Lambda\eta_{ij} = kT_{ij}
\ea
Using the expression for $r(l)$ and $\psi (l)=1$ 
we get the nontrivial field equations as
\ba
\4{{b_0}^2}{({b_0}^2 + l^2)^2} + \Lambda &=& k V(T)(1+ T^{\p 2})^{1/2} \label{00}\\
-\4{b_0^2}{(b_0^2 + l^2)^2} + \Lambda &=& k \4{V(T)}{(1+ T^{\p 2})^{1/2}} \label{11}
\ea
From (\ref{00}) and (\ref{11}) we get after dividing and multiplying both sides of the equations, respectively,
\ba
1+T^{\p 2} = \4{\Lambda + \4{{b_0}^2}{({b_0}^2 + l^2)^2}}{\Lambda - \4{{b_0}^2}
{({b_0}^2 + l^2)^2}} \la{teqn}\\
k^2 V^2 = \Lambda^2-\4{b_0^4}{(b_0^2 + l^2)^4} \la{veqn}
\ea
The last equation constrains the values of $\Lambda$: either $\Lambda\ge b_0^{-2}$ 
or $\Lambda \le -b_0^{-2}$ so that the function $V(T)$ is real for all 
possible values of $l$. 
(In fact the first condition also follows from use of the Jacobian elliptic 
integral to solve the equation (\ref{teqn}) as we show below.) 
In the rest of the calculations we assume $\Lambda > b_0^{-2}$. It can be 
easily shown that the negative value of $\Lambda$ gives rise to complex valued 
tachyon field. Solving equation (\ref{teqn}) we obtain the tachyon field 
in terms of the Jacobian Elliptic Integral of the first kind 
(denoted by $F({\mathrm {amplitude,modulus}})$ \cite{byrd,abra,elliptic}) as follows: 
\begin{widetext}
\ba
T(l)=(\4{2b_0}{\Lambda^{1/2}(1+\Lambda^{1/2}b_0)})^{\4{1}{2}}{\mathrm {F}}
(\tan^{-1}(\4{l}{\sqrt{b_0^2 - \4{b_0}{\Lambda^{1/2}}}}),
(\4{2}{1+b_0\Lambda^{1/2}})^{1/2}) \label{tsol}\\ 
l=(b_0^2 - \4{b_0}{\Lambda^{1/2}})^{\4{1}{2}}\tan({\mathrm{am}}(\sqrt{\4{\Lambda^{1/2}
(1+b_0 \Lambda^{1/2})}{2b_0}}T,\sqrt{\4{2}{1+b_0 \Lambda^{1/2}}}) \label{lsol}
\ea
\end{widetext}
The function $V(T)$ can be obtained from (\ref{veqn}) after replacing $l$  
in $V(l)$ in terms of $T$ field (\ref{lsol}). The field in this case has 
limiting values $ \pm \pi /2$ as $l \rightarrow \pm \infty$ respectively. 
So, asymptotically, this function takes constant values equal in magnitude, 
opposite in sign, as the modulus remains  constant. Also the function under 
consideration is odd under change of sign of the argument. Thus the field has a kink-like configuration with spherical symmetry. 
A plot of the field as a function of the proper radial coordinate is shown in Fig \ref{fieldplot}. The plot for the potential as function of the field is shown 
in Fig \ref{potplot}. The values of the two constants for both the plots are 
$\Lambda=4, b_0 =1$ and we assume $k=1$. We use arbitrary units in these plots.

One can ask what would happen if we chose the second non--constant 
red-shift function in the presence of a positive or negative cosmological constant. 
The answer is that we do not have a real tachyon field for all values of $l$ and secondly, the resulting redshift function has a zero for some real $l$ making the choice unacceptable in both cases.

So far we have obtained an exact solution of the Einstein field equations for 
the case of the tachyon field (depending on radial coordinate only) coupled 
to gravity in a static spherically symmetric spacetime of the specific 
form (\ref{metric}) with $\psi^2(l)$ taken to be constant. The course of work 
showed that 
we need to introduce the $ positive \,\, \Lambda$ term which, effectively, modifies 
the stress-energy tensor of the system. This was crucial and the solution 
exists only when $\Lambda > b_0^{-2}$. The energy density of the tachyon field is given by 
\be
\rho(T) = \4{1}{k}(G_{00} - \Lambda) = -(\4{r^{\p \p}}{r} + \Lambda)
\ee
and is negative definite in the range of $\Lambda$ as specified (for all values 
of $l$). Fig 3 shows a plot of the energy density as a function of the
field $T$.  Note that the energy density is negative but bounded below. 
The energy conditions are violated, as expected, by this solution. As can be seen from (\ref{Tt},\ref{Tr},\ref{Tth}) the radial and transverse pressures are positive definite. 
\begin{figure}[tbh!]
\begin{minipage}[t]{5cm}
\includegraphics[width=0.9\textwidth]{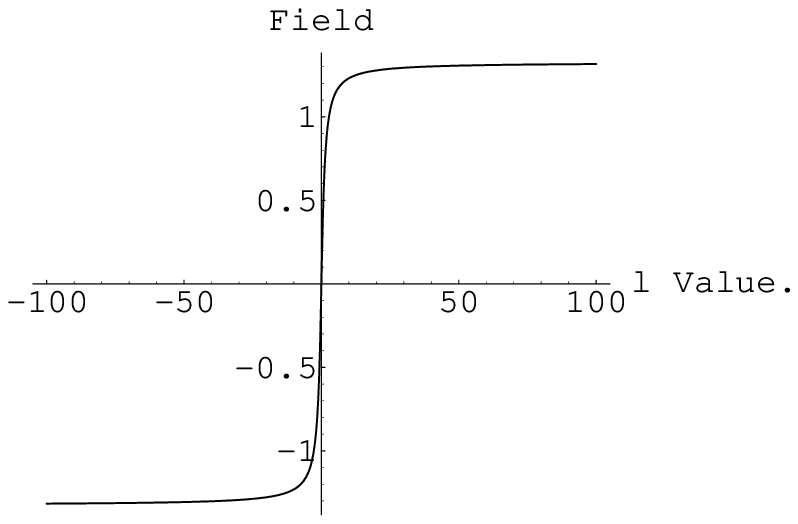} \caption{Variation of tachyon field with proper radial distance.} \label{fieldplot}
\end{minipage}
\hfill
\begin{minipage}[t]{5cm}
\includegraphics[width=0.9\textwidth]{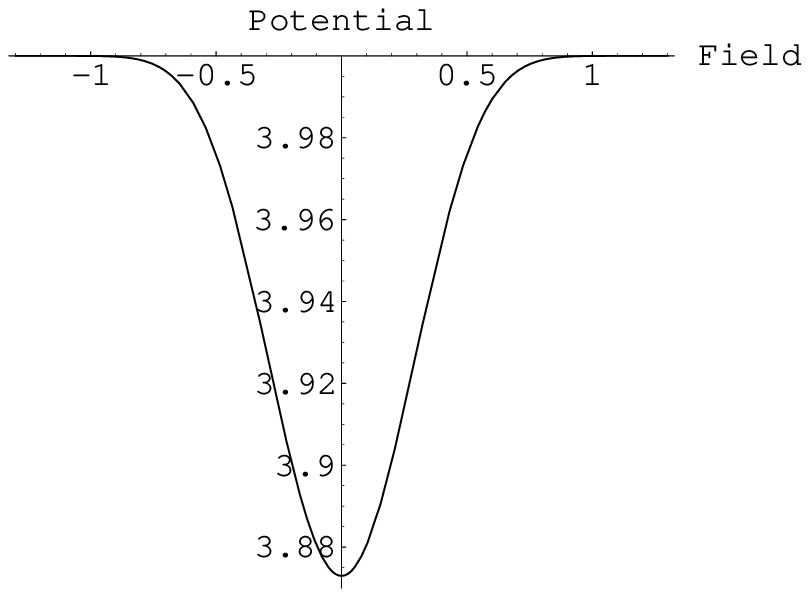} \caption{Variation of potential with field.} \label{potplot}
\end{minipage}
\hfill
\begin{minipage}[t]{5cm}
\includegraphics[width=0.9\textwidth]{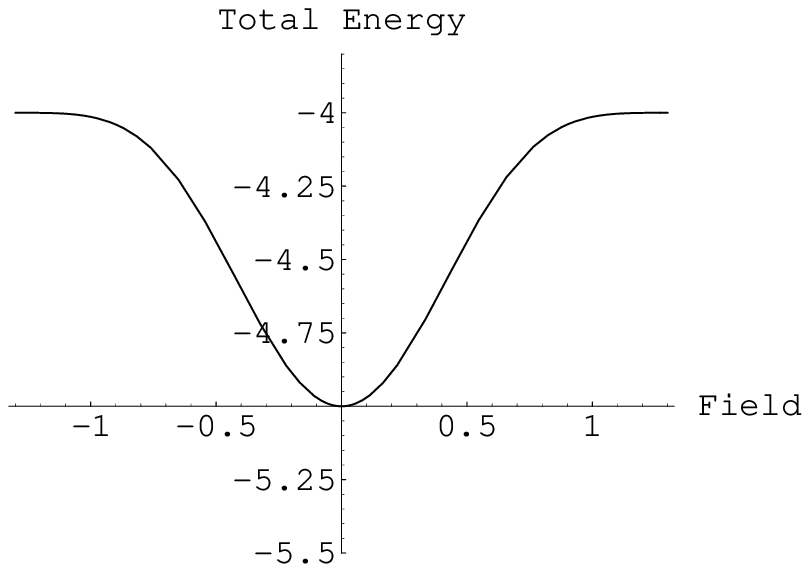} \caption{Variation of total energy density with  field.} \label{energyplot}
\end{minipage}
\end{figure}

\noindent {\sf Linear Stability : } It is useful to test the stability of 
the tachyon field configuration and the metric under small perturbations. 
We assume linear, time independent, scalar type metric as well as field 
perturbations of the form:
\ba
r(l)\rightarrow {\bar r}(l)&=&r(l)+\delta r(l), \,\, (\delta r(l) << r(l))\\
T(l)\rightarrow {\bar T}(l)&=&T(l)+ \delta T(l), \,\, (\delta T(l) << T(l))
\ea
For simplicity the metric perturbation is confined to the angular part only. 
We neglect all square and higher order terms in $\delta r(l)/r(l)$ and 
$\delta T(l)/T(l)$ and their products in our calculations. The perturbed form of eqn(\ref{cons}) with $\psi=1$ is 
\be
(\delta r)^{\p \p} + 2\frac{r^{\p}}{r}(\delta r)^{\p}+\4{r^{\p\p}}{r}(\5r)=0
\ee 
Using the perturbed Einstein equation, $\delta G_{\mu \nu}=k \, \delta T_{\mu \nu}$ for 
the $(tt)$ and $(\theta\theta)$ components we derive the perturbation equation 
for the field as follows 
\be
V^{\p}(T)(1+T^{\p2})(2+T^{\p2})\5T(l)+V(T)T^{\p3}\5T^{\p}=0
\ee
which simplifies to
\be
\4{(\5T)^{\p}(l)}{(\5T)(l)}+\frac{4b_0 l \Lambda (b_0^2+l^2)}{\sqrt{2}(\Lambda(b_0^2+l^2)^2 - b_0^2)^{3/2}}  =0
\ee
The solutions for the $\5r(l)$ and $\5T(l)$ equations are
\ba
\5r(l)&=&\4{c_1l}{(1+l^2/b_0^2)^{1/2}} + \4{c_2}{(1+l^2/b_0^2)^{1/2}} \label{dreqn}\\
\5T(l)&=&c_3\exp \left[b_0\left({\frac{2}{\Lambda(b_0^2+l^2)^2-b_0^2}}\right)^{1/2}\right]
\ea
where $c_1$, $c_2$ and $c_3$ are integration constants.  
{\it Thus linear perturbations in both $r(l)$ and $T(l)$ 
are finite everywhere}. The tachyon field perturbation is maximum at the throat 
and decays with increasing proper radial distance. In the case of $\5r(l)$, we notice the presence of even and odd parity solutions 
both of which are finite everywhere and also as $l\rightarrow \inf$. 
Since the metric and field perturbations do not diverge in the entire spacetime, we conclude that our solution is stable against the above restricted type of small, linear, scalar perturbations.

\noindent {\sf $d=2+1$ and $d>4$ dimensions :} 
We will conclude by making some observations 
on wormhole solutions corresponding to the action (\ref{action})
in dimensions other than $d=4$. In $(2+1)$ dimensions, using the following 
static, circularly symmetric line element                                                                                                   
\be
ds^2=-\chi^2(l)dt^2+dl^2+r^2(l)d\theta^2
\ee
and the corresponding action in the same dimension, the relation $G_{tt}=-G_{\theta\theta}$ imply 
\be
 \4{\chi^{\p\p}(l)}{\chi^3(l)}=\4{r^{\p\p}(l)}{r^3(l)}\implies \chi(l)=\pm r(l)
\ee
The resulting nontrivial Einstein equations without the $\Lambda$ term for either of the signs in the last relation are
\ba
\4{r^{\p\p}(l)}{r(l)}&=&kV(T)(1+T^{\p 2})^{1/2}\\
\4{r^{\p 2}(l)}{r^2(l)}&=&\4{kV(T)}{(1+T^{\p 2})^{1/2}}
\ea
We are once more forced to choose one of the three unknown functions 
involved here. The analogue of the Ellis geometry here would lead to the 
ansatz: $r(l)=(b_0^2+l^2)^{1/2}$ which however gives rise to a solution 
valid only in certain finite domain of $l$ and is discarded. Addition 
of a $\Lambda$ term can not solve the problem. So we try a different ansatz, 
$r(l)= \chi(l)= \cosh(\4{l}{b_0}+\beta)$ where $b_0$ and $\beta$ are  two constants. This gives consistent solutions as follows:
\ba
T(l)&=&b_0\log|\tanh(\4{1}{2}(\4{l}{b_0}+\beta))|\\
V(T)&=&\4{1}{kb_0^2}\tanh(2\tanh^{-1}(e^{T/b_0}))
\ea

A point to note is that the above solution contains a nontrivial redshift 
function whose presence is a matter of necessity in $(2+1)$ dimension
(without a $\chi (l)$ we obtain only flat spacetime as solution to the
field equations with a tachyon matter source). Both the redshift function 
and the shape function have desirable wormhole properties. The constant $\beta$ is introduced to make the field well behaved at $l=0$.
Considering the coordinate $l$ as an `extra dimension' one might imagine
the $t, \theta$ part of the line element as a toy 1--brane with $R\times
S^1$ topology. The overall factor of this $l=constant$ $1+1$ dimensional 
section is the so--called `warp factor' {\cite{braneworld}}.

In dimensions $d>4$ it turns out that the $r$ coordinate is more 
convenient to work with than the proper radial distance $l$. 
Assuming a $d-2$ sphere instead of the 2--sphere in $d=4$ we
write down the line element as :

\be
ds^2=-dt^2 + \4{dr^2}{1-\4{b(r)}{r}}+r^2d\Omega_{d-2}^2
\ee

Notice that we have disregarded the redshift function here for
simplicity.

The $(tt)$ and ($jj$) ($j$ being any angular coordinate) components of the energy momentum tensor
(for tachyon--matter source) 
imply once again $G_{tt}=-G_{jj}$ which leads to the relation
\ba
(d-3)\4{b(r)}{r^3}&=&-\4{1}{2r^3}(rb^{\p}(r)-b(r))\\
\mathrm{so \,\,that\,\,\,\,} \,\4{b(r)}{r}&=&Cr^{2(3-d)}
\ea
$C$ being an integration constant. This can be regarded as a 
wormhole solution by choosing proper value of $C$ for a fixed $d>4$. 
(Notice that at the throat, $b(r_0)=r_0$.) The solution is 
asymptotically flat and the proper radial distance is well behaved everywhere.  Thus it is possible to obtain wormhole solutions corresponding to the 
action (\ref{action}) both in lower and in higher dimensions with different 
characteristics. The solutions in higher and lower dimensions are however not 
`Ellis' wormholes but can be thought of as its counterparts in some
restricted sense.

\noindent{\sf Concluding remarks :} Let us summarize briefly what we
have achieved. It was known from the work of Ellis and others that
the negative kinetic energy phantom can be a source for the 
Ellis geometry in $3+1$ dimensions. In this article, we have shown that
the tachyon condensate can be a source for Ellis geometry. In fact, the
spatially dependent tachyon field configuration is kink--like. Recall that  
wormhole solutions supported by a
kink type scalar field are known in the literature {\cite{sergey}}.
However, the so--called tachyon kinks in flat spacetime 
obtained in the literature are all singular in nature {\cite{sen2}}.
The kink solution we obtain has a negative energy density but 
it is a solution in a curved spacetime background. It is not
clear to us whether such a solution has any relevance in the
context of string theory and D-branes. As mentioned before, even without a 
direct motivation from string theory, one might
imagine our solution as a way of obtaining wormholes using
non-standard matter fields (and actions) as sources. In addition to
obtaining the solution we have also found that it is stable against
linear perturbations. We have also briefly explored extensions in $2+1$ 
and $D>4$ dimensions.

Several unanswered questions do exist. Firstly, if the phantom scalar
and the tachyon scalar give the same geometry is there is a relation
between them atleast in a class of background spacetimes? Further,
in $D=4$ if we assume a wormhole shape function different from Ellis
geometry, then, is there a way in which the tachyon matter can still be
a source via the choice of a proper red--shift function? In $D>4$
what happens if we include contributions from the Gauss--Bonnet term 
{\cite{sk2}} and/or
include a redshift function in the line element? We hope to return to
some of these issues in the near future.

Note added: We mention here that the article by H. Shinkai and S. A.
Hayward, 2002 {\it Phys. Rev. D} {\bf 66} 044005 discusses the dynamic
stability of Ellis wormholes in considerable detail. Additionally, linear
stability of Ellis wormholes (with different scalar sources) have also been
discussed by Bronnikov K A, Cl\'{e}ment G, Constantinidis C P \& Fabris J C,
1998 {\it Phys. Lett.} {\bf A243}, 121 and  Armendariz-Picon C, 2002 {\it Phys.
Rev. D} {\bf 65} 104010, the latter being in the realm of scalar tensor theory.

\begin{acknowledgments}
AD is thankful to the Council of Scientific and Industrial Research, India for grant of a Junior Research Fellowship.
\end{acknowledgments}


\begin{thebibliography}{99}
\bibitem{elliptic} For definitions and properties of the Jacobian Elliptic Integrals, consult {\cite{byrd, abra}}. We give here the necessary definitions:
The Jacobian Elliptic integral of the first kind is defined by :
$F(\phi,k)=\int^{\phi}_0 \4{d\theta}{\sqrt{1-k\sin^2 \theta}}$.
Here $\phi$ and $k$ are known respectively as  the amplitude and
the modulus of the elliptic integral. The inverse relation is given by:
$\phi=\mathrm{JacobiAmplitude}(F,k)$ or simply $\mathrm{am}(F,k)$. We use,
in particular, the following result $\int_0^y \4{dt}{\sqrt{(t^2+a^2)(t^2+b^2)}} = \4{1}{a}F(\phi,k)$
where $\phi = \tan^{-1}\4{y}{b}; \, k^2=\4{a^2-b^2}{a^2};\, a>b;\, y>0$.
Although stated strictly for $y>0$ in result 221.00 of \cite{byrd} we have extended it for the
case $y \leq 0$ also using the fact that the integrand is an even function of
the argument. Notice that the condition $a>b$ in the present case gives rise to the condition $\Lambda > b_0^{-2}$.
\end{thebibliography}
\end{document}